\documentclass[nofootinbib]{revtex4-1}
\usepackage{silence}
\usepackage{bm}
\usepackage{hyperref}
\usepackage{amsmath,amssymb,amsfonts,graphicx,slashed,bbm}
\usepackage[paper=letterpaper,margin=1in]{geometry}
\usepackage{braket, color}
\usepackage{upgreek}

\newcommand{\beq}{\begin{eqnarray}}
\newcommand{\eeq}{\end{eqnarray}}
\usepackage{amsmath}
\usepackage{amssymb}
\usepackage[paper=letterpaper,margin=1in]{geometry}
\usepackage{braket}
\usepackage{upgreek}
\usepackage{tikz}

\begin{document}

\title{ Free at Last: Bose Metal Uncaged}
\author{Philip W. Phillips}

\affiliation{Department of Physics and Institute for Condensed Matter Theory,
University of Illinois
1110 W. Green Street, Urbana, IL 61801, U.S.A.}

\date{\today}

\begin{abstract}
A Bose metallic phase disrupts the classic two-state superconductor-insulator picture.

\end{abstract}
\maketitle

Even the particle world is not immune to identity politics. Bosons have been in a bit of an identity crisis, or so it has seemed since 1989\cite{phillips}. Quantum mechanics requires bosons made of two paired electrons either to condense into a superfluid with a well-defined phase (yellow region in Fig. ({\ref{data}) in which all the purple arrows point in the same direction) with zero electrical resistance or localize in an insulating state with infinite resistance.   The direct transition from superconducting to insulating states was widely observed in a range of thin films\cite{goldman,hebard,kapm}.  The most popular model for explaining these observations\cite{fisher}  claims that the destruction of superconductivity occurs when the resistance of the thin film exceeds a critical value.  Namely, if it is greater than the quantum of resistance, $R_q = h/4e^2$,  for bosons on the brink
of localization, insulating behaviour obtains, else superconductivity persists.   Here $h$ is Planck's constant and $e$ is the electric charge.  On p. 1505 of this issue, Yang et al.\cite{Yang1505} {\it offer} a counterexample by establishing that a bosonic metallic phase disrupts the superconductor-insulator transition (SIT) in the high-temperature superconductor YBCO.

Indeed, bosons have been toying with metallicity since 1989\cite{goldman}. However, the older data were dismissed as being due to a refrigeration problem.  The explanation was that the bosons were not cooling, remaining in suspended animation and conducting. Nonetheless, the metallic states were observed at a frequency that kept hope alive for a true zero Kelvin disruption of the SIT.  In the magnetic-field tuned transition in MoGe\cite{kapm}, the resistance flattened at an onset temperature that decreased with increasing field. If the metal were due to a heating effect, then the failure to cool the sample would occur at the same temperature, not one determined by the magnetic field\cite{kapm}. Similar observations were made on Ta\cite{yoon} and in single crystals of ion-gated ZrNCl\cite{saito}.

A direct attack on the refrigeration problem came from the experiments that showed the metallic state in amorphous InO$_x$\cite{hebard} and crystalline 2H-NbSe$_2$\cite{tsen} were artifacts\cite{shahar} as they vanished by eliminating all sources of electromagnetic radiation. Yang, et al.\cite{Yang1505}employed  this now standard filtering scheme to directly address what is carrying the current in the metallic phase.

\begin{figure}
\includegraphics[scale=0.5]{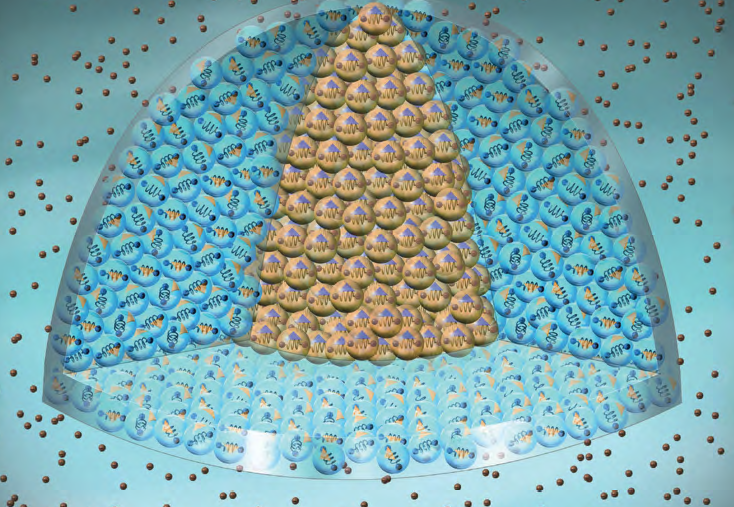} 
  \caption{Superconducting state shown in yellow has bosons composed of two paired electrons. The phase of the Cooper pairs is indicated by the red arrow.  In the superconducting state, all the arrows point in the same direction indicative of phase coherence in a superconducting state.  This paper shows that the metallic state (in blue) is bosonic (paired electrons) with no particular orientation of the phase.   Ultimately the electron pairs break apart liberating single-electron excitations beyond the glassy metallic phase.} \label{data}
\end{figure}

The experiment\cite{Yang1505} utilized a series of high-temperature superconducting YBCO films, which were tuned through the superconductor-insulator transition by reactive-ion-etching (RIE). This technique transfers a patterned array of hexagonally plated holes to YBCO.   The authors observed that the normal state thin film resistance increases with increasing etching time.   The etching time controls the thickness of the resulting pores'' sidewalls, straining the YBCO shell and  creating structural defects. The defects increase the resistance of the thin film sheet. The transition the authors observed is an example of disorder-tuned SIT and similar to the 1989 experiments\cite{goldman}.

 The authors\cite{Yang1505} observe three phases with a critical value for the resistance roughly twice the quantum of resistance, $R_q$\cite{fisher}.   The metallic phase persists down to 50mK. To characterize this phase, the authors confirmed previous reports\cite{brezany} of a vanishing  Hall resistance. This type of response to a magnetic field is an indication of emergent particle-hole symmetry. The novelty in the authors'' experiments is their measurement of the resistance oscillations as a function of field.  In a typical metal in which electrons are the charge carriers, the oscillations have a period of $h/e$.   Here instead, the authors find an $h/2e$ periodicity indicative of charge $2e$ carriers.  This remarkable experiment thereby proves definitively that bosons are free to roam in a metallic phase invalidating the standard picture of the two-phase model\cite{fisher}. 
 
Hence, the teasing is over. Bosons can exist as a metal. Further, the authors\cite{Yang1505} observe the amplitude of the conductance oscillations grows in the superconducting state, remains constant in the metal, and decreases in the insulator. This trend is consistent with a divergent phase coherent length in the superconductor, a finite size in the Bose metal (see blue region in Fig. (\ref{data}), and no phase coherence in the insulator. 

The logically correct starting point for the Bose metal is an array of superconducting islands connected by random tunneling matrix elements.  While adding an underlying metallic matrix to this model can achieve the metallic state\cite{kivelson},  this is not consistent with the experiments because the charge carriers have charge $2e$ not $e$.   Because of the disorder, glassy bosonic models are a natural choice. In fact, sate-of-the-art numerical simulations have shown that randomness
in superconducting island models in two dimensions produces a glassy phase with a superfluid stiffness\cite{stiffness}
that scales as $L^{-0.39}$ ($L^{0.27}$in three-dimensions) (L the system size) and hence vanishes in the thermodynamics limit in 2D.
Such glassy models have a non-zero bosonic conductivity[14] with emergent particle-hole symmetry and
hence contain the requisite ingredients needed to explain the disorder-induced Bose metallic state in the
experiments reported here.   Such glassy models (blue region in Fig. (\ref{data})) have a non-zero bosonic conductivity\cite{phillips,maymann} with emergent particle-hole symmetry and hence contain the requisite ingredients needed to explain the disorder-induced Bose metallic state in the experiments reported here.   

Experiments similar to those reported here should be carried out on the magnetic-field tuned transition to discern if the charge carriers
are 2e as well.  If proven to be the case, then the Bose metal represents the general exception to the standard paradigm\cite{fisher}.  In addition, time-resolved measurements could reveal the low-energy excitations of the newly
confirmed Bose metal, whose true novelty is metallicity without the fermionic organizing principle of Pauli exclusion.

\acknowledgements Funding  from  NSF DMR19-19143 is acknowledged.

%
\end{document}